\documentclass[twocolumn,english,aps,preprintnumbers,amsmath,amssymb,superscriptaddress]{revtex4}
\usepackage[latin9]{inputenc}
\usepackage{amsmath}
\usepackage{graphicx}

\makeatletter
\@ifundefined{textcolor}{}
{%
 \definecolor{BLACK}{gray}{0}
 \definecolor{WHITE}{gray}{1}
 \definecolor{RED}{rgb}{1,0,0}
 \definecolor{GREEN}{rgb}{0,1,0}
 \definecolor{BLUE}{rgb}{0,0,1}
 \definecolor{CYAN}{cmyk}{1,0,0,0}
 \definecolor{MAGENTA}{cmyk}{0,1,0,0}
 \definecolor{YELLOW}{cmyk}{0,0,1,0}
}

\usepackage{dcolumn}% Align table columns on decimal point
\usepackage{bm}% bold math
\usepackage{color}
\usepackage[final]{pdfpages}

\makeatother

\usepackage{babel}

\begin{document}

\preprint{preprint(\today)}

\title{Strong coupling nature of kagome superconductivity in LaRu$_{3}$Si$_{2}$}

\author{C.~Mielke III}
\thanks{These authors contributed equally to the paper.}
\affiliation{Laboratory for Muon Spin Spectroscopy, Paul Scherrer Institute, CH-5232
Villigen PSI, Switzerland}
\affiliation{Physik-Institut, Universitat Z\"{u}rich, Winterthurerstrasse 190, CH-8057 Zurich, Switzerland}

\author{Y.~Qin}
\thanks{These authors contributed equally to the paper.}
\affiliation{Wuhan National High Magnetic Field Center and School of Physics, Huazhong University of Science and Technology, Wuhan 430074, China}

\author{J.-X.~Yin}
\thanks{These authors contributed equally to the paper.}
\affiliation{Laboratory for Topological Quantum Matter and Spectroscopy, Department of Physics, Princeton University, Princeton, NJ, 08544, USA}

\author{H.~Nakamura}
\thanks{These authors contributed equally to the paper.}
\affiliation{Institute for Solid State Physics, University of Tokyo, Kashiwa, 277-8581, Japan}

\author{D.~Das}
\affiliation{Laboratory for Muon Spin Spectroscopy, Paul Scherrer Institute, CH-5232
Villigen PSI, Switzerland}

\author{K.~Guo}
\affiliation{International Center for Quantum Materials and School of Physics, Peking University, Beijing, China}
\affiliation{CAS Center for Excellence in Topological Quantum Computation, University of Chinese Academy of Science, Beijing, China}

\author{R.~Khasanov}
\affiliation{Laboratory for Muon Spin Spectroscopy, Paul Scherrer Institute, CH-5232
Villigen PSI, Switzerland}

\author{J. Chang}
\affiliation{Physik-Institut, Universitat Z\"{u}rich, Winterthurerstrasse 190, CH-8057 Zurich, Switzerland}

\author{Z.Q. Wang}
\affiliation{Department of Physics, Boston College, Chestnut Hill, MA, USA}

\author{S.~Jia}
\affiliation{International Center for Quantum Materials and School of Physics, Peking University, Beijing, China}
\affiliation{CAS Center for Excellence in Topological Quantum Computation, University of Chinese Academy of Science, Beijing, China}

\author{S.~Nakatsuji}
\affiliation{Institute for Solid State Physics, University of Tokyo, Kashiwa, 277-8581, Japan}

\author{A.~Amato}
\affiliation{Laboratory for Muon Spin Spectroscopy, Paul Scherrer Institute, CH-5232
Villigen PSI, Switzerland}

\author{H.~Luetkens}
\affiliation{Laboratory for Muon Spin Spectroscopy, Paul Scherrer Institute, CH-5232
Villigen PSI, Switzerland}

\author{G.~Xu}
\email{gangxu@hust.edu.cn} 
\affiliation{Wuhan National High Magnetic Field Center and School of Physics, Huazhong University of Science and Technology, Wuhan 430074, China}

\author{Z.M. Hasan}
\email{mzhasan@princeton.edu} 
\affiliation{Laboratory for Topological Quantum Matter and Spectroscopy, Department of Physics, Princeton University, Princeton, NJ, 08544, USA}

\author{Z.~Guguchia}
\email{zurab.guguchia@psi.ch} 
\affiliation{Laboratory for Muon Spin Spectroscopy, Paul Scherrer Institute, CH-5232 Villigen PSI, Switzerland}

%\pacs{74.20.Mn, 74.25.Ha, 74.70.Xa, 76.75.+i}

\begin{abstract}

%Electronic systems with flat bands can be a fertile ground for hosting emergent phenomena including unconventional magnetism and superconductivity. Flat band superconductivity has recently been discovered in twisted bilayer graphene with a phase diagram similar to high-temperature superconductivity. The flat bands can also naturally arise in geometrically frustrated lattice systems including kagome lattices. However, a flat band kagome superconductor remains elusive. 
We report muon spin rotation (${\mu}$SR) experiments together with first-principles calculations on microscopic properties of superconductivity in the kagome superconductor LaRu$_{3}$Si$_{2}$ with $T_{\rm c}$ ${\simeq}$ 7K. We find that the calculated normal state band structure features a kagome flat band and Dirac as well as van Hove points formed by the Ru-$dz^{2}$ orbitals near the Fermi level. Below $T_{\rm c}$, ${\mu}$SR reveals isotropic type-II superconductivity, which is robust against hydrostatic pressure up to 2 GPa. Intriguingly, the ratio  2$\Delta/k_{\rm B}T_{\rm c}$ ${\simeq}$ 4.3 (where ${\Delta}$ is the superconducting energy gap) is in the strong coupling limit, and $T_{\rm c}$/$\lambda_{eff}^{-2}$ (where ${\lambda}$ is the penetration depth) is comparable to that of high-temperature unconventional superconductors. We also find that electron-phonon coupling alone can only reproduce small fraction of $T_{\rm c}$ from calculations, which suggests other factors in enhancing $T_{\rm c}$ such as the correlation effect from the kagome flat band, the van Hove point on the kagome lattice, and high density of states from narrow kagome bands. Our experiments and calculations taken together point to strong coupling and the unconventional nature of kagome superconductivity in LaRu$_{3}$Si$_{2}$. 

\end{abstract}

\maketitle

\newpage

%%%%%%%%%%%%%%%%%%%%%%%%%%%%%%%%%%%%%%%%%%%%%%%%%%%%%%%%%%%%%%
\begin{figure*}[t!]
\centering
\includegraphics[width=0.8\linewidth]{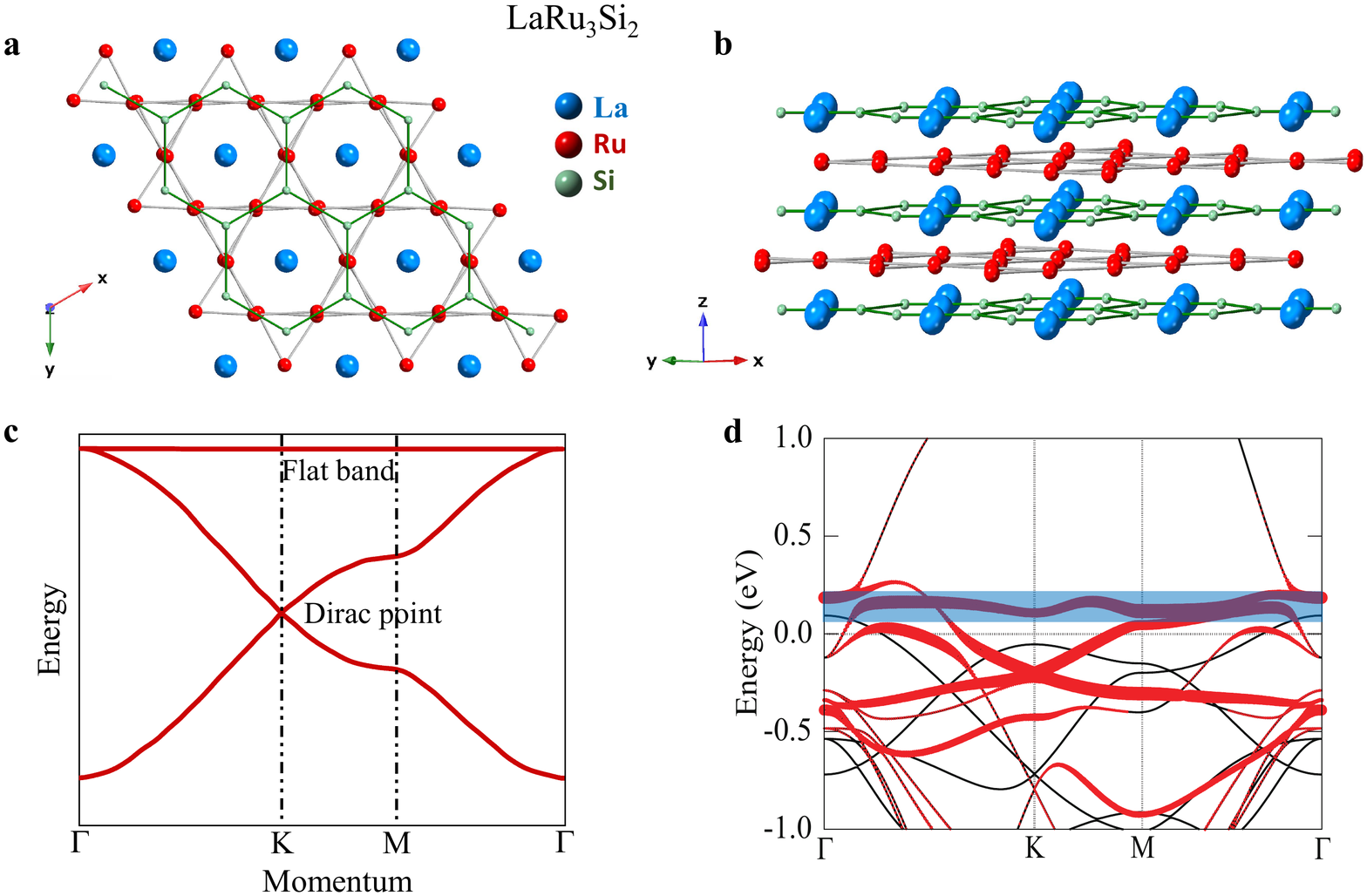}
\vspace{-0.8cm}
\caption{ (Color online)
Top view (a) and side view (b) of the atomic structure of LaRu$_{3}$Si$_{2}$. The Ru atoms construct a kagome lattice (red middle size circles), while the Si (green small size circles) and La atoms (blue large size circles) form a honeycomb and triangular structure, respectively. (c) Tight-binding band structure of kagome lattice exhibiting two Dirac bands at the $K$-point and a flat band across the whole Brillouin zone. (d) The band structures (black) and orbital-projected band structure (red) for the Ru-$dz^{2}$ orbital without SOC along the high symmetry $k$-path, presented in conformal kagome BZ. The width of the line indicates the weight of each component. The blue-colored region highlights the manifestation of the kagome flat band.}
\label{fig2}
\end{figure*}
%%%%%%%%%%%%%%%%%%%%%%%%%%%%%%%%%%%%%%%%%%%%%%%%%%%%%%%%%%%%%%

%\section{Introduction}

  Layered systems with highly anisotropic electronic properties have been found to be potential hosts for rich, unconventional and exotic quantum states. The prominent class of layered materials is the kagome-lattice systems \cite{JXYin2,LYe,THan,JXYin1,GuguchiaCSS,Yazyev,Mazin}. The kagome lattice refers to a two-dimensional network of corner-sharing triangles. Crystalline materials that contain a kagome lattice have attracted considerable attention because of the associated electronic band structure and frustrated antiferromagnetic (AFM) interactions. This band structure reveals a pair of Dirac points similar to those found in graphene, and a dispersionless, flat band that originates from the kinetic frustration associated with the geometry of the kagome lattice. Flat bands are exciting because the associated high density of electronic states hints at possible correlated electronic phases when found close to the Fermi level \cite{Hofmann2020,Nunes,Volovik}. In general, it has been conjectured that the presence of flat bands in a two-dimensional system may give rise to room temperature superconductivity \cite{Volovik,Heikkila,Kopnin}. The possibility to access flat bands and their influence on the physical properties of the system has been studied for about three decades \cite{Lieb,Tanaka,Noda,Julku,Hartman}. Recently, superconductivity was discovered in bilayer graphene \cite{CaoPablo} when its individual layers are twisted with respect to each other by a specific angle giving rise to a flat band. The general understanding is that, due to the presence of flat bands, the kinetic energy is quenched and interaction-driven quantum phases prevail. Recent Monte Carlo calculations on a two-dimensional system \cite{Hofmann2020}, focusing on the case with topological flat bands, demonstrate that the ground state is a superconductor and find a broad pseudogap regime that exhibits strong pairing fluctuations and even a tendency towards electronic phase separation. Moreover, a square-octagon lattice was theoretically studied in which two perfectly flat bands were found \cite{Nunes} and the calculated superconducting phase diagram was found to have two superconducting domes \cite{Nunes}, as observed in several types of unconventional superconductors. Thus, there is a resurgence of interest in flat bands as a means to explore unconventional superconductivity from the experimental front \cite{Hofmann2020,Peotta,Sayyad}.

 In this framework, the  layered system LaRu$_{3}$Si$_{2}$ \cite{Barz,Vandenberg,Kishimoto,LiWen,LiZeng} appears to be a good example of a material hosting both a kagome lattice and superconductivity. The structure of LaRu$_{3}$Si$_{2}$ contains distorted kagome layers of Ru sandwiched between layers of La and layers of Si having a honeycomb structure (see Fig. 1a and b), crystallizing in the $P$6$_{3}$/m space group (see Supplementary Figure S4). The system was shown to be a typical type II superconductor with a SC transition temperature with the onset as high as ${\simeq}$ 7 K \cite{Kishimoto}. Anomalous properties \cite{LiZeng} in the normal and SC states \cite{Kishimoto} were reported in LaRu$_{3}$Si$_{2}$, such as the deviation of the normal state specific heat from the Debye model, non-mean field like suppression of superconductivity with magnetic field and non-linear field dependence of the induced quasiparticle density of states (DOS). However, for the most part only the critical temperatures and fields have been characterized for the superconducting state of LaRu$_{3}$Si$_{2}$. Thus, thorough and microscopic exploration of superconductivity in LaRu$_{3}$Si$_{2}$ from both experimental and theoretical perspectives are required in order to understand the origin of the relatively high value of the critical temperature. 
 
Here, we combine powerful microscopic probe such as the muon spin rotation (${\mu}$SR) \cite{Sonier,Brandt,GuguchiaNature,LukeTRS} together with first-principles calculations to elucidate the superconductivity in kagome superconductor LaRu$_{3}$Si$_{2}$ with $T_{\rm c}$ ${\simeq}$ 7 K. We find that the calculated normal state band structure features a kagome flat band, Dirac point and van Hove point formed by the Ru-$dz^{2}$ orbitals near the Fermi level. In the superconducting state, the $T_{\rm c}$/$\lambda_{eff}^{-2}$  ratio is comparable to those of unconventional superconductors. The relatively high $T_{\rm c}$ for the low carrier density may hint at an unconventional pairing mechanism in LaRu$_{3}$Si$_{2}$. Moreover, the measured SC gap value $\Delta$ = 1.14(1) meV yields a BCS ratio 2$\Delta/k_{\rm B}T_{\rm c}$ ${\simeq}$ 4.3, suggesting that the superconductor LaRu$_{3}$Si$_{2}$ is in the strong coupling limit. We also find that the electron-phonon coupling alone turns out to be insufficient to reproduce the experimental critical temperature $T_{\rm c}$, suggesting other factors in enhancing $T_{\rm c}$.

%  such as the correlation effect from the kagome flat band and the high DOS from the group of narrow kagome bands.  

%\section{Results}
%\subsection{Kagome lattice band structure for LaRu$_{3}$Si$_{2}$}
%\subsection{Crystal and electronic structure and superconducting transition}
 Figures 1a and b show the top view and side view, respectively, of the atomic structure of LaRu$_{3}$Si$_{2}$. The Ru atoms construct a distorted kagome lattice, while the Si and La atoms form honeycomb and triangular structure, respectively. Calculations show that the lattice constant and phonon dispersion is very sensitive to the exchange correlation potential and the value of the Hubbard $U$. A phonon dispersion with negative frequency modes is obtained when only the GGA or LDA type of exchange correlation potential is used (See supplementary Fig. S9) \cite{Kresse,Perdew,Anisimov,Gianozzi,Togo}. In the GGA + Hubbard $U$ framework, the negative frequency modes can be eliminated when U ${\textgreater}$ 1 eV. However, the lattice constant $c$ shows a sudden collapse when U increases minutely. Moreover, we notice that the electronic structure is not sensitive to the exchange correlation functional.  All these results indicate that Coulomb repulsion $U$ plays a crucial role in stabilizing the crystal structure of LaRu$_{3}$Si$_{2}$, indicating the medium strength correlations in this system. This is reminiscent of another family of correlated metals, the parent compounds of iron-based superconductors, in which GGA/LDA is also unable to describe bond length and bond strength, but can describe band structures quantitively well.

%Calculations show that Coulomb repulsion $U$ plays a crucial role in stabilizing the crystal structure of LaRu$_{3}$Si$_{2}$ (see supplementary information), indicating the medium correlations in this system. This is reminiscent of another family of correlated metals, the parent compounds of iron-based superconductors, in which GGA/LDA is also unable to describe bond length and bond strength, but can describe band structures quantitively well. 

The calculated total and projected density of state (DOS) (supplementary figure S7a), demonstrate that the states at the Fermi level in LaRu$_{3}$Si$_{2}$ are mainly contributed by the Ru 4$d$ electrons. The band structure with the Ru-$dz^{2}$ orbital projection is shown in Figure 1d. There are several bands that cross the Fermi level and the complex three dimensional Fermi surfaces (see the supplementary figure S7c) are formed in the first Brillouin zone, indicating multi-band physics. Most importantly, in the $k_{z}$ = 0 plane, a flat band of the kagome lattice formed by the Ru-$dz^{2}$ orbitals is found 0.1 eV above the Fermi level, highlighted by blue-colored region in Figure 1d. In addition, a Dirac point at the K (K$^{`}$)-point with linear dispersion is found 0.2 eV below the Fermi level. Moreover, the van Hove point on the kagome lattice at M point can be clearly seen in Fig. 1d, which is located even closer to the Fermi energy (${\sim}$ 50 meV). Thus, the system LaRu$_{3}$Si$_{2}$ exhibits, around the Fermi level, a typical kagome lattice band structure (see Figure 1c), revealing a Dirac point, the van Hove point and a dispersionless, flat band that originates from the kinetic frustration associated with the geometry of the kagome lattice. 

%\subsection{Probing the superconducting vortex state as a function of temperature and pressure}

%%%%%%%%%%%%%%%%%%%%%%%%%%%%%%%%%%%%%%%%%%%%%%%%%%%%%%%%%%%%%%
\begin{figure*}[t!]
\centering
\includegraphics[width=1.0\linewidth]{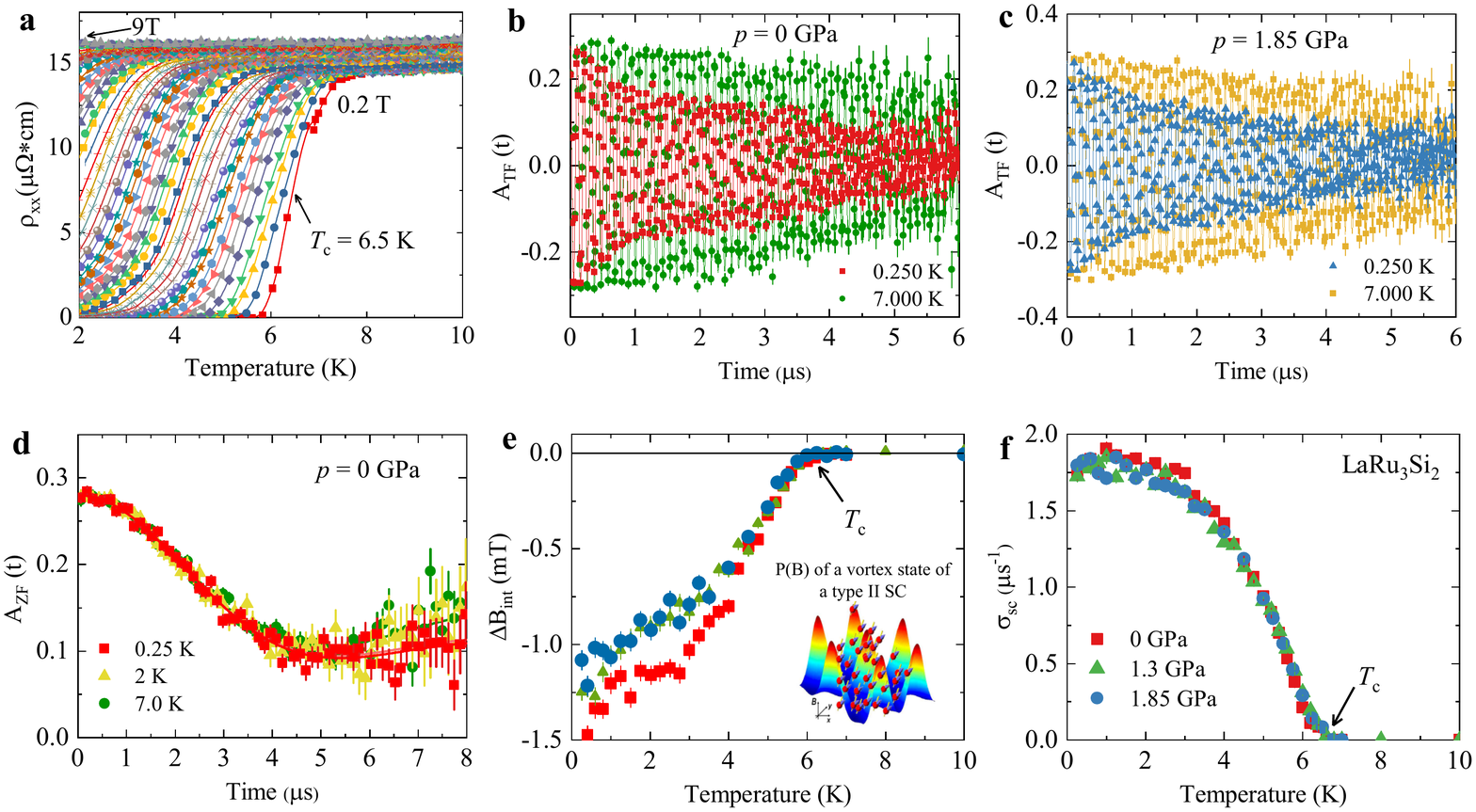}
\vspace{-3.0cm}
\caption{ Color online)
(a) The temperature dependence of the electrical resistivity for LaRu$_{3}$Si$_{2}$, recorded for various applied magnetic fields. (b-d) Transverse-field (TF) and zero-field (ZF) ${\mu}$SR time spectra for LaRu$_{3}$Si$_{2}$ to probe the SC vortex state and magnetic responses, respectively. The TF spectra are obtained above and below $T_{\rm c}$ (after field cooling the sample from above $T_{\rm c}$): (b)  $p$ = 0 GPa and (c) $p$ = 1.85 GPa. The solid lines in panels a and b represent fits to the data by means of Eq.~1. The dashed lines are a guide to the eye. Inset illustrates how muons, as local probes, sense the inhomogeneous field distribution in the vortex state of type-II superconductor. (d) ZF ${\mu}$SR time spectra for LaRu$_{3}$Si$_{2}$ recorded above and below $T_{\rm c}$. The line represents the fit to the data using a standard Kubo-Toyabe depolarization function \cite{Toyabe}, reflecting the field distribution at the muon site created by the nuclear moments. Temperature dependence of the diamagnetic shift ${\Delta}$$B_{\rm dia}$ (e) and the muon spin depolarization rate ${\sigma}_{\rm sc}$($T$) (f), measured at various hydrostatic pressures in an applied magnetic field of ${\mu}_{\rm 0}H = 70$~mT. The arrows mark the $T_{c}$ values.}
\label{fig2}
\end{figure*}
%%%%%%%%%%%%%%%%%%%%%%%%%%%%%%%%%%%%%%%%%%%%%%%%%%%%%%%%%%%%% 

%%%%%%%%%%%%%%%%%%%%%%%%%%%%%%%%%%%%%%%%%%%%%%%%%%%%%%%%%%%%%%%
\begin{figure*}[t!]
\centering
\includegraphics[width=0.8\linewidth]{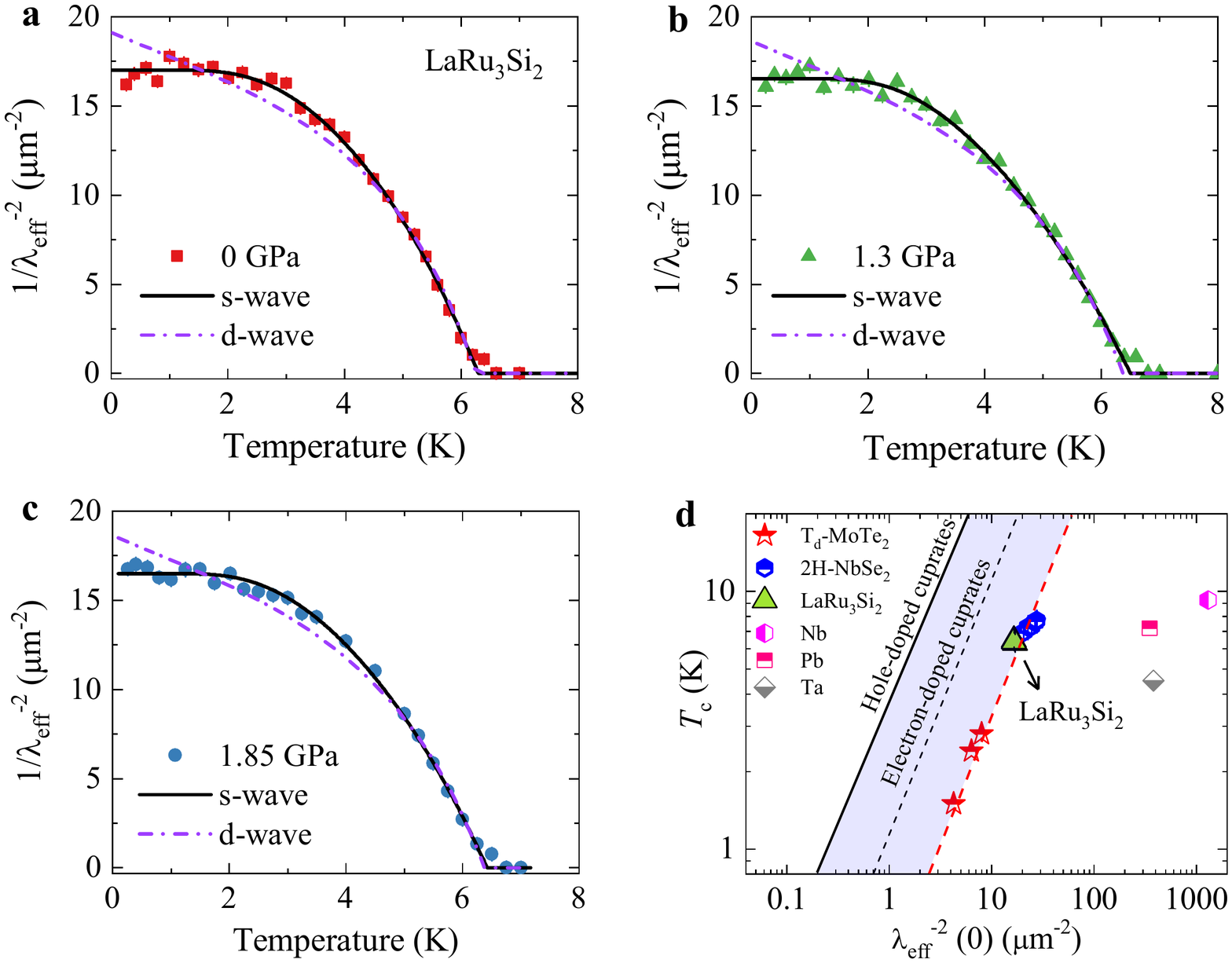}
\vspace{-0.8cm}
\caption{ (Color online)  
The temperature dependence of ${\lambda}_{eff}^{-2}$ measured at various applied hydrostatic pressures for LaRu$_{3}$Si$_{2}$: (a) $p$ = 0 GPa, (b) 1.3 GPa, and (c) 1.85 GPa. The solid line corresponds to a $s$-wave model and the dashed line represents fitting with a $d$-wave model. (d) A plot of $T_{\rm c}$ versus the ${\lambda}_{eff}^{-2}(0)$ obtained from our ${\mu}$SR experiments in LaRu$_{3}$Si$_{2}$. The dashed red line represents the relation obtained for layered transition metal dichalcogenide superconductors $T_{d}$-MoTe$_{2}$ \cite{GuguchiaMoTe2} and 2H-NbSe$_{2}$ \cite{GuguchiaNbSe2}. The data are taken at ambient as well as under pressure. The relation observed for underdoped cuprates is also shown (solid line for hole doping \cite{Uemura1,Uemura3,Uemura4,Uemura5} and the dashed black line for electron doping \cite{Shengelaya,Luetkens}). The points for various conventional BCS superconductors are also shown.}
\label{fig4}
\end{figure*}
%%%%%%%%%%%%%%%%%%%%%%%%%%%%%%%%%%%%%%%%  

  The temperature dependence of electrical resistivity for LaRu$_{3}$Si$_{2}$, depicted in Figure 2a under different applied fields up to 9 T, shows superconductivity with the onset and the midpoint (at 50 ${\%}$ drop of the resistivity) of the transition at 7 K and 6.5 K, respectively. The critical magnetic field was estimated to be as high as ${\mu}_{\rm 0}$$H_{\rm cr}$ = 8.5 T at $T$ = 2 K (see the supplementary Figure S6). In the following, we provide the microscopic details of the superconductivity in this system using ${\mu}$SR.

 Figures 2b and c exhibit the transverse field (TF) ${\mu}$SR time spectra for LaRu$_{3}$Si$_{2}$ in an applied magnetic field of 70 mT, measured at $p$ = 0 GPa and maximum applied pressure $p$ = 1.85 GPa, respectively. The spectra above (7 K) and below (0.25 K) the SC transition temperature $T_{{\rm c}}$ are shown. Above $T_{{\rm c}}$, the oscillations show a small damping due to the random local fields from the nuclear magnetic moments. Below $T_{{\rm c}}$ the damping rate strongly increases with decreasing temperature due to the presence of a nonuniform local magnetic field distribution as a result of the formation of a flux-line lattice (FLL) in the SC state. Magnetism, if present in the samples, may enhance the muon spin depolarization rate and falsify the interpretation of the TF-${\mu}$SR results. Therefore, we have carried out zero-field (ZF)-${\mu}$SR experiments above and below $T_{{\rm c}}$ to search for magnetism (static or weakly fluctuating) in LaRu$_{3}$Si$_{2}$. As shown in Fig. 2d, no sign of either static or fluctuating magnetism could be detected in ZF time spectra down to 0.25 K. The spectra are well described by a Kubo-Toyabe depolarization function \cite{Toyabe,GuguchiaPressure}, reflecting the field distribution at the muon site created by the nuclear moments of the sample and the pressure cell. 
%Moreover, no change in ZF-${\mu}$SR relaxation rate (see Fig. 2d) across $T_{c}$ was observed, pointing to the absence of any spontaneous magnetic fields associated with a TRS \cite{LukeTRS,HillierTRS,BiswasTRS} breaking pairing state in LaRu$_{3}$Si$_{2}$. 
 Returning to the discussion of the TF-${\mu}$SR data, we observe a strong diamagnetic shift of the internal magnetic field ${\mu}_{0}$$H_{\rm int}$ sensed by the muons below $T_{{\rm c}}$. This is evident in Figure 2e, where we plot the temperature dependence of ${\Delta}$$B_{\rm dia}$ = ${\mu}_{0}$($H_{\rm int,SC}$ - $H_{\rm int,NS}$), i.e., the difference between the internal field ${\mu}_{0}$$H_{\rm int,SC}$ measured in the SC fraction and ${\mu}_{0}$$H_{\rm int,NS}$ measured in the normal state at $T$ = 10 K. The strong diamagnetic shift excludes the occurrence of field induced magnetism in LaRu$_{3}$Si$_{2}$. The absence of magnetism in zero-field or under applied magnetic fields implies that the increase of the TF relaxation rate below $T_{{\rm c}}$ is solely arising from the FLL in the superconducting state.

From the TF-${\mu}$SR time spectra, we determined the Gaussian superconducting relaxation rate ${\sigma}_{{\rm sc}}$ (after subtracting the nuclear contribution), which is proportional to the second moment of the field distribution (see Method section). Figure 2f shows ${\sigma}_{\rm sc}$ as a function of temperature for LaRu$_{3}$Si$_{2}$ at ${\mu}_{\rm 0}H=0.07$~T, recorded for various hydrostatic pressures. Below $T_{\rm c}$ the relaxation rate ${\sigma}_{\rm sc}$ starts to increase from zero due to the formation of the FLL. We note that both ${\sigma}_{\rm sc}$ and $T_{\rm c}$ stay nearly unchanged under pressure, indicating a robust superconducting state of LaRu$_{3}$Si$_{2}$.
At all pressures, ${\sigma}_{\rm sc}(T)$ shows saturation towards low temperatures. We show in the following paragraph that the observed temperature dependence of ${\sigma}_{{\rm sc}}$, which reflects the topology of a SC gap, is consistent with the presence of the single $s$-wave SC gap on the Fermi surface of LaRu$_{3}$Si$_{2}$. 
 
%\subsection{Superconducting pairing symmetry} 
 
  In order to investigate the symmetry of the SC gap, we note that temperature dependence of the magnetic penetration depth ${\lambda}(T)$ is related to the relaxation rate ${\sigma}_{{\rm sc}}(T)$, in the presence of a perfect triangular vortex lattice by the equation \cite{Brandt}: 
\begin{equation}
\frac{\sigma_{sc}(T)}{\gamma_{\mu}}=0.06091\frac{\Phi_{0}}{\lambda_{eff}^{2}(T)},
\end{equation}
where ${\gamma_{\mu}}$ is the gyromagnetic ratio of the muon, and
${\Phi}_{{\rm 0}}$ is the magnetic-flux quantum. Thus, the flat $T$-dependence
of ${\sigma}_{{\rm sc}}$ observed at various pressures for low temperatures
(see Fig.~2f) is consistent with a nodeless superconductor, in which
$\lambda_{eff}^{-2}\left(T\right)$ reaches its zero-temperature value exponentially. We note that it is the effective penetration depth $\lambda_{eff}^{-2}$ (powder average), which we extract from the ${\mu}$SR depolarization rate (Eq. (1)), and this is the one shown in the figures. The magnetic penetration depth is one of the fundamental parameters of a superconductor, since it is related to the superfluid density $n_{s}$ via 1/${\lambda}_{eff}^{2}$ = $\mu_{0}$$e^{2}$$n_{s}/m^{*}$ (where $m^{*}$ is the effective mass and $n_{s}$ is the SC carrier density). A quantitative analysis of ${\lambda}_{eff}(T)$ \cite{Bastian,Tinkham,carrington,Sonier} is described in the methods section of the supplementary information. The results of this analysis are presented in Fig. 3a-c, where the temperature dependence of $\lambda_{eff}^{-2}$ for LaRu$_{3}$Si$_{2}$ is plotted at various pressures. The solid and dashed lines represent fits to the data using $s$-wave and $d$-wave models, respectively. As seen in Fig. 3, ${\lambda}_{eff}$($T$) dependence is best described by a momentum independent $s$-wave model with a gap value of $\Delta$ = 1.2(1)~meV and $T_{\rm c}$ ${\simeq}$ 6.5~K. The effective penetration depth, $\lambda_{eff}$, at zero temperature is found to be 240(10)~nm. We note that the ($p_{\rm x}$ + i$p_{\rm y}$) pairing symmetry is also characterised by a gap function without nodes
 in 2D systems and would also result in saturated behaviour at low temperatures. However, the possibility of $p_{\rm x}$ + i$p_{\rm y}$ pairing is excluded by the absence of a time-reversal symmetry (TRS) breaking state. Also a $d$-wave gap symmetry was tested, but was found to be inconsistent with the data (see the dashed line in Fig~3a-c). In particular, it is difficult to account for the very weak temperature dependence of ${\lambda}_{eff}$($T$) at low-$T$ within models  possessing nodes in the gap function. We also tested the power law (1-(T/$T_{c}$)$^{2}$) which has been proposed theoretically \cite{Hirshfeld} for the superfluid density of dirty $d$-wave superconductors and found it to be inconsistent with the data. This leaves a nodeless or fully gapped state as the most plausible bulk SC pairing state in the bulk of LaRu$_{3}$Si$_{2}$. The observed single gap superconductivity in this multi-band system implies that the superconducting pairing involves predominately one band. However, if the inter-band coupling is strong than one may detect single-gap like behavior of the superfluid density in multigap materials \cite{Gupta}.

%The ratio of the SC gap to $T_{\rm c}$ was estimated to be (2$\Delta/k_{\rm B}T_{\rm c}$) ${\simeq}$ 4.3. Thus, the ratio $T_{\rm c}$/$\lambda_{eff}^{-2}$ ${\sim}$ 0.4 indicates the relatively high $T_{\rm c}$ for a small number of carriers.

%The effective penetration depth, $\lambda_{eff}$, at zero temperature is found to be 240(10)~nm.  In general, the penetration depth ${\lambda}$ is given as a function of $n_{\rm s}$, $m^{*}$, ${\xi}$ and the mean free path $l$ as
% 
%\begin{equation}
%\begin{aligned}
%\frac{1}{\lambda^2} = \frac{4\pi n_se^2}{m^*c^2} \times  \frac{1}{1 + \xi/l}, 
%%\label{eq3}
%\end{aligned}
%\end{equation}  
% 
%For systems close to the clean limit, ${\xi}$/$l$ ${\rightarrow}$ 0, the second term essentially becomes unity, and the simple relation 1/${\lambda}^{2}$ $\propto$ $n_{s}/m^{*}$ holds. Considering the $H_{c2}$ value of LaRu$_{3}$Si$_{2}$ \cite{Kishimoto}, we estimated ${\xi}$ ${\simeq}$ 16 nm for $p$ = 0 GPa respectively.  Thus, in view of the short coherence length, we can assume that LaRu$_{3}$Si$_{2}$ lies close to the clean limit \cite{Frandsen}. With this assumption, we obtain the ground-state value  $n_{s}/(m^{*}/m_{e}$) ${\simeq}$ 2.8 ${\times}$ 10$^{26}$ m$^{-3}$. Interestingly, $n_{s}/(m^{*}/m_{e}$) is robust with respect to applied hydrostatic pressure.  

%\subsection{Pairing strength and the superfluid density}

 The ratio of the superconducting gap to $T_{\rm c}$ for LaRu$_{3}$Si$_{2}$ was estimated to be (2$\Delta/k_{\rm B}T_{\rm c}$) ${\simeq}$ 4.3, which is consistent with the strong coupling limit BCS expectation \cite{GuguchiaMoTe2}. However, a similar ratio can also be expected for Bose Einstein Condensate (BEC)-like picture as pointed out in ref. \cite{Uemura4}. To place this system LaRu$_{3}$Si$_{2}$ in the context of other superconductors, in Fig. 3d we plot the critical temperature $T_{\rm c}$ against the superfluid density $\lambda_{eff}^{-2}$. Most unconventional superconductors have $T_{\rm c}$/$\lambda_{eff}^{-2}$ values of about 0.1-20, whereas all of the conventional BCS superconductors lie on the far right in the plot, with much smaller ratios. In the other words, unconventional superconductors are characterized by a dilute superfluid (low density of Cooper pairs) while conventional BCS superconductors exhibit dense superfluid.
Moreover, a linear relationship between $T_{\rm c}$ and $\lambda_{eff}^{-2}$ is expected only on the Bose Einstein Condensate (BEC)-like side and is considered a hallmark feature of unconventional superconductivity. We recently showed that the linear correlation is an intrinsic property of the superconductivity in transition metal dichalcogenides \cite{GuguchiaMoTe2,GuguchiaNbSe2}, whereas the ratio $T_{\rm c}$/$\lambda_{eff}^{-2}$ is lower than the ratio observed in hole-doped cuprates (see Figure 3d). For twisted bilayer graphene \cite{CaoPablo} the ratio $T_{\rm c}$/$\lambda_{eff}^{-2}$ was found to be even higher than the one for cuprates. For LaRu$_{3}$Si$_{2}$, the ratio is estimated to be $T_{\rm c}$/$\lambda_{eff}^{-2}$ ${\simeq}$ 0.37, which is approximately a factor of 15 lower than the one for hole-doped cuprates, but still being far away from conventional BCS superconductors. Interestingly, the point for LaRu$_{3}$Si$_{2}$ is almost perfectly located on the trend line on which charge density wave superconductors 2H-NbSe$_{2}$ and 4H-NbSe$_{2}$ as well as Weyl-superconductor $T_{d}$-MoTe$_{2}$ \cite{GuguchiaMoTe2} lie, as shown in Fig. 3d. Both systems 2H-NbSe$_{2}$ and $T_{d}$-MoTe$_{2}$ are also strong coupling superconductors, as we previously reported. This finding hints at an unconventional pairing mechanism in LaRu$_{3}$Si$_{2}$ with a low density of Cooper pairs and similar electron correlations as in 2H-NbSe$_{2}$ and $T_{d}$-MoTe$_{2}$, but much weaker electron correlations than in cuprates and twisted bilayer graphene.

%\section{Discussion}

  Since the present muon spin rotation experiments show direct evidence for the absence of local moments of the Ru atom in LaRu$_{3}$Si$_{2}$, this kagome system is different from high-$T_{\rm c}$ superconductors or spin liquid compounds. It is rather on the side of itinerant kagome or hexagonal systems such as (Cs,K)V$_{3}$Sb$_{5}$ \cite{Eric} or even 2H-NbSe$_{2}$ \cite{GuguchiaNbSe2}, where ${\mu}$SR shows the absence of magnetic correlations. On the other hand, the system LaRu$_{3}$Si$_{2}$ does not exhibit a CDW ground state unlike the superconductors (Cs,K)V$_{3}$Sb$_{5}$ \cite{JiaxinKVSb}. One possibility is that this system is just somewhat away from CDW order and have higher optimal $T_{\rm c}$. The fact that the stability of the crystal structure is obtained with the addition of the Hubbard $U$ suggests proximity of this superconductor LaRu$_{3}$Si$_{2}$ to a CDW, since $U$ opposes the CDW. Pressure independent superfluid density was recently reported for CDW free layered transition metal dichalcogenide system 2M-WS$_{2}$ \cite{GuguchiaWS2}, while a large enhancement of $\lambda_{eff}^{-2}$ was found in CDW superconductor 2H-NbSe$_{2}$ under pressure when suppressing the CDW order \cite{GuguchiaNbSe2}. Thus, the robustness of both $T_{\rm c}$ and the superfluid density $\lambda_{eff}^{-2}$ of LaRu$_{3}$Si$_{2}$ to hydrostatic pressure strongly suggests that $T_{\rm c}$ has the optimal value already at ambient pressure and that the system is away from a competing CDW ground state. 
        
In order to understand the origin of $T_{\rm c}$, we compare experimentally measured critical temperatures with those calculated using the McMillan equation. The phonon dispersion for LaRu$_{3}$Si$_{2}$, calculated by the GGA+U method with $U$ = 1 eV, is shown in Supplementary Figure S7d. It consists of only the positive frequency modes, and the optimized lattice constants agree very well with the experimental parameters. Based on the phonon dispersion, we calculated the electron-phonon coupling constant ${\lambda}$ to be ${\sim}$ 0.45. Then by using the Debye temperature ${\theta}$ = 280 K and the Coulomb pseudo potential ${\mu}^{*}$ = 0.12, the superconducting transition temperature $T_{\rm c}$ was estimated. The electron-phonon coupling induced critical temperature was found to be $T_{\rm c}$ ${\sim}$ 1 K, which is obviously smaller than the experimental value. Since the electron-phonon coupling can reproduce only  some fraction of the experimental $T_{\rm c}$, other factors in enhancing $T_{\rm c}$ must be considered. The calculations show the presence of a flat band near the Fermi level, which may enhance correlations and can contribute to the enhancement of $T_{\rm c}$ in this system. However, the flat band is 100 meV above $E_{\rm f}$ and it may not have a key role in enhancing $T_{\rm c}$. The van Hove point on the kagome lattice at M point, which can be seen in Fig. 1d, is located even closer to the Fermi energy (${\sim}$ 50 meV), which can also contribute to the enhancement of $T_{\rm c}$. This van Hove point at M is of a similar distance to $E_{\rm f}$ (below $E_{\rm f}$) in KV$_{3}$Sb$_{5}$, and is what was shown to drive the 2x2 CDW order \cite{JiaxinKVSb} at much higher temperatures than $T_{\rm c}$. Moreover, we find that the whole kagome bands are somewhat narrow (${\sim}$ 300 meV), which may also enhance $T_{\rm c}$ through the overall higher density of states (see the supplementary information). The narrowness of the kagome bands to be ${\sim}$ 300 meV may be formally similar to a group of narrow bands, found in twisted bilayer graphene \cite{CaoPablo}. 

%Finally, relatively high $T_{\rm c}$ in LaRu$_{3}$Si$_{2}$ is attributed to the combination of electron-phonon coupling, correlations from flat band, the van Hove point on the kagome lattice at M point and high DOS from narrow kagome bands. 

%Additional work is needed to obtain a microscopic understanding of the connection between kagome band electrons and superconductivity.

%\section{Summary}

In summary, we provide the first microscopic investigation of superconductivity in the layered distorted kagome superconductor LaRu$_{3}$Si$_{2}$ with a bulk probe. Specifically, the zero-temperature magnetic penetration depth ${\lambda}_{eff}\left(0\right)$ and the temperature dependence of ${\lambda_{eff}^{-2}}$ were studied by means of ${\mu}$SR experiments. The superfluid density is best described by the scenario of a gap function without nodes. Interestingly, the $T_{\rm c}$/$\lambda_{eff}^{-2}$ ratio is comparable to those of high-temperature unconventional superconductors, pointing to the unconventional nature of superconductivity in LaRu$_{3}$Si$_{2}$. Moreover, the measured SC gap value yields a BCS ratio 2$\Delta/k_{\rm B}T_{\rm c}$ ${\simeq}$ 4.3, suggesting that the superconductor LaRu$_{3}$Si$_{2}$ is in the strong coupling limit. Furthermore, we find the calculated normal state band structure features a kagome flat band, Dirac point and van Hove point formed by the Ru-$dz^{2}$ orbitals near the Fermi level. The electron-phonon coupling induced critical temperature $T_{\rm c}$, estimated from the phonon dispersion, was found to be smaller than the experimental value. Thus, the enhancement of $T_{\rm c}$ in LaRu$_{3}$Si$_{2}$ is attributed to the presence of the flat band and the van Hove point relatively close to the Fermi level as well as to the high density of states from the narrow kagome bands. Our experiments and calculations taken together point to strong coupling and the unconventional nature of kagome superconductivity in LaRu$_{3}$Si$_{2}$. The present results have implications for ongoing investigations of LaRu$_{3}$Si$_{2}$ and other kagome lattice superconductors and will stimulate theoretical studies to obtain a microscopic understanding of the connection between kagome band electrons and superconductivity.\\

%This suggests other factors in enhancing $T_{\rm c}$ such as the correlation effect from the kagome flat band, the van Hove point on the kagome lattice and the high density of states from the group of narrow kagome bands.

%%%%%%%%%%%%%%%%%
\section{Acknowledgments}~
The ${\mu}$SR experiments were carried out at the Swiss Muon Source (S${\mu}$S) Paul Scherrer Insitute, Villigen, Switzerland. Z. Guguchia thanks Rafael Fernandes for useful discussions. M.Z.H. acknowledges visiting scientist support from IQIM at the California Institute of Technology. Z.Q.W. is supported by DOE grant No. DE-FG02-99ER45747.
G. Xu and Y. Qin would like to thank the support by the National Key Research and Development Program of China (2018YFA0307000) and the National Natural Science Foundation of China (11874022).

%\textbf{\section{Contributions}}
%Project planning:  Z.G.; Sample growth: F.v.R. and R.J.C.;
%${\mu}$SR experiments:  Z.G.; Z.S.; R.K.; A.A.; H.L.; C.B.; E.M; A.S.; B.F., Z.G.; and Y.J.U.;
%${\mu}$SR data analysis: Z.G.; Data interpretation: Z.G., A.R.W., A.N.P. and Y.J.U.;
%X-ray pair distribution function measurements: S.B., Z.G., and S.B.;
%DFT calculations: A.T.L., and C.A.M.; 
%Draft writing: Z.G., with contributions and/or comments from all authors. 

%\includepdf[pages={-}]{MoTe2Supplementary.pdf}
%\includepdf[pages={3}]{MoTe2Supplementary.pdf}
%\includepdf[pages={4}]{MoTe2Supplementary.pdf}

\newpage

\renewcommand{\figurename}{Supplementary Figure S}
\maketitle
\section{Supplementary Information}

\textbf{General remark}: Here, we concentrate on muon spin rotation/relaxation/resonance  ($\mu$SR) \cite{Sonier,Brandt,GuguchiaNature} measurements of the magnetic penetration depth $\lambda$ in LaRu$_{3}$Si$_{2}$, which is one of the fundamental parameters of a superconductor, since it is related to the superfluid density $n_{s}$ via 1/${\lambda}^{2}$ = $\mu_{0}$$e^{2}$$n_{s}/m^{*}$ (where $m^{*}$ is the effective mass). Most importantly, the temperature dependence of ${\lambda}$ is particularly sensitive to the structure of the SC gap. Moreover, zero-field ${\mu}$SR is a very powerful tool for detecting static or dynamic magnetism in exotic superconductors, because very small internal magnetic fields are detected in measurements without applying external magnZ.etic fields. In addition, the band structure and phonon spectrum of LaRu$_{3}$Si$_{2}$ were also investigated through DFT calculations.\\ 

\textbf{Sample preparation}: Polycrystalline samples of LaRu$_{3}$Si$_{2}$ were synthesized by melting the mixture of La, Ru, and Si ($\textgreater$ 3N) in Ar atmosphere using an arc-melting method. To prevent the formation of LaRu$_{2}$Si$_{2}$, 10 ${\%}$ excess of Ru was added in the starting composition. The synthesized samples were characterized by the powder X-ray diffraction method and subsequent Rietveld analyses (RIETAN-FP \cite{Izumi}). The samples were confirmed to be LaRu$_{3}$Si$_{2}$ with about 5 wt${\%}$ of Ru impurity.\\

We performed the multi-phase Rietveld analyses for LaRu$_{3}$Si$_{2}$ with $P$6$_{3}$/$m$ symmetry \cite{Vandenberg}, where Ru atoms form a distorted Kagome-lattice, and for small amount of Ru impurity (see supplementary Figure S4). For both phases, the XRD results agree well with the simulated patterns. Since the crystal structure with $P$6/$mmm$ symmetry, where Ru atoms form a perfect Kagome-lattice, is also reported \cite{Godart}, we examined the actual crystal structure of the LaRu$_{3}$Si$_{2}$ phase in the samples. In supplementary Figure S4, the peak positions of the two crystal structures and our XRD patterns were compared. The peaks for (2-13), (3-11) and (3-21) planes, which should be absent in $P$6/$mmm$ symmetry, were observed in our XRD measurement. This indicates that the crystal structure of the LaRu$_{3}$Si$_{2}$ samples has $P$6$_{3}$/$m$ symmetry.\\

\textbf{Superconducting transition}: The temperature dependences of zero-field cooled and field-cooled dc magnetic susceptibility of LaRu$_{3}$Si$_{2}$, measured in 1~mT, are shown in Figure S5.  The data evidence sharp superconducting onset at $T_{\rm c}$ ${\simeq}$ 6.5 K, which is in agreement with the critical temperature obtained from microscopic ${\mu}$SR method and the midpoint of the SC transition from the resistivity measurements. The value of ZFC susceptibility at 2 K corresponds to 100 ${\%}$ volume fraction of superconductivity. The temperature dependence of the critical magnetic field for LaRu$_{3}$Si$_{2}$, estimated from resistivity measurements is shown in Figure S6. The critical magnetic field was estimated to be as high as ${\mu}_{\rm 0}$$H_{\rm cr}$ = 8.5 T at $T$ = 2 K.\\ 

%%%%%%%%%%%%%%%%%%%%%%%%%%%%%%%%%%%%%%%%%%%%%%%%%%%%%%%%%%%%%
\begin{figure*}[b!]
\centering
\includegraphics[width=0.7\linewidth]{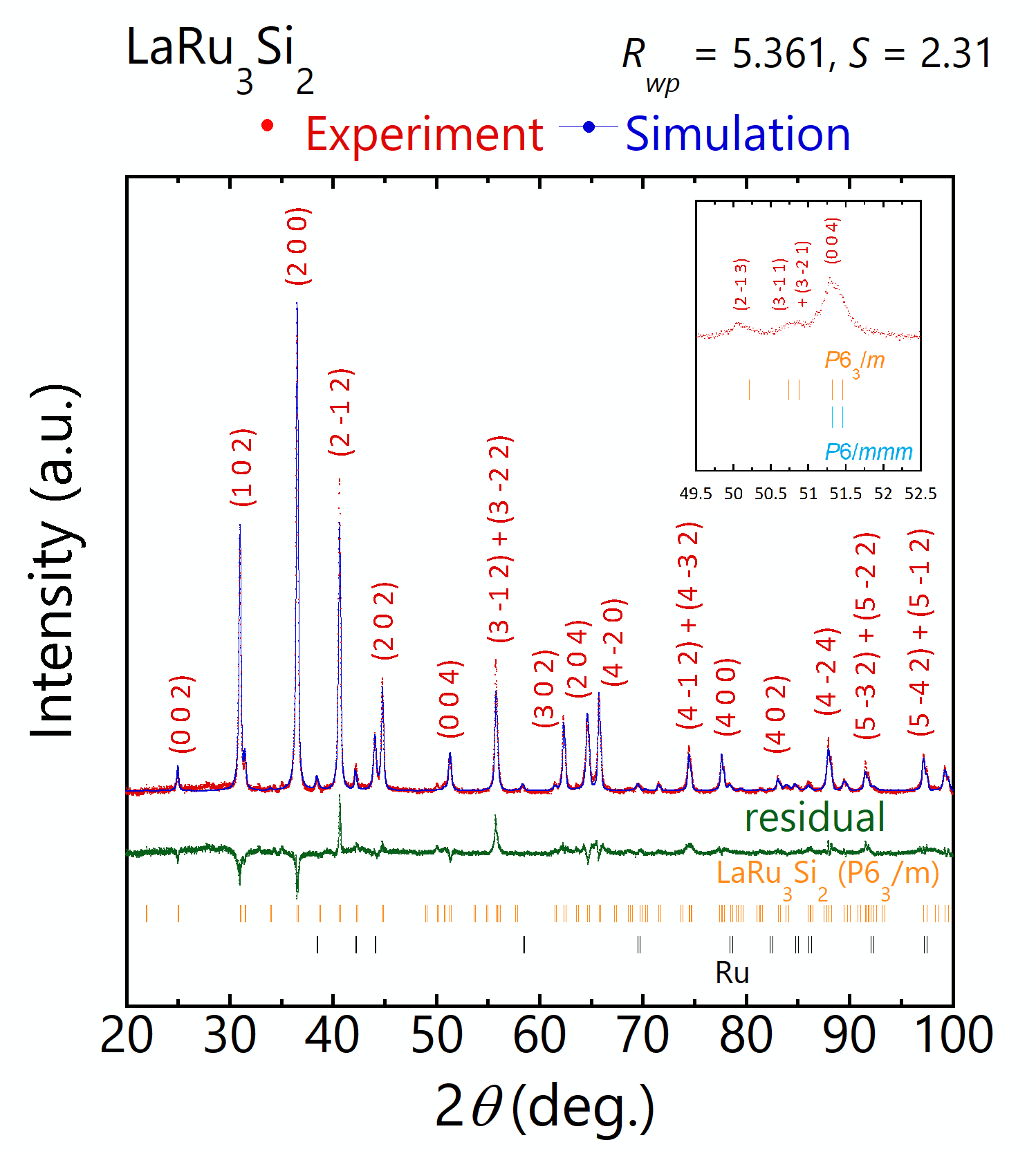}
\vspace{0cm}
\caption{ (Color online) Powder X-ray (Cu-$K_{\alpha}$) diffraction patterns of LaRu$_{3}$Si$_{2}$. The red circles and blue solid lines represent the experimental and simulation patterns of Rietveld analysis, respectively. The residual errors of the simulation pattern (green solid line), the peak positions for LaRu$_{3}$Si$_{2}$ (orange bar) and Ru (black bar), and the major peak indexes (red texts) are shown. Inset: Experimental XRD pattern around 2${\theta}$ = 51$^{\deg}$. The peak positions in $P$6$_{3}$/$m$ and $P$6/$mmm$ symmetries are shown with orange and light-blue bar, respectively.}
\label{fig4}
\end{figure*}
%%%%%%%%%%%%%%%%%%%%%%%%%%%%%%%%%%%%%%%%%%%%%%%%%%%%%%%%%%%%%%   

%%%%%%%%%%%%%%%%%%%%%%%%%%%%%%%%%%%%%%%%%%%%%%%%%%%%
\begin{figure*}[htb!]
\includegraphics[width=0.7\linewidth]{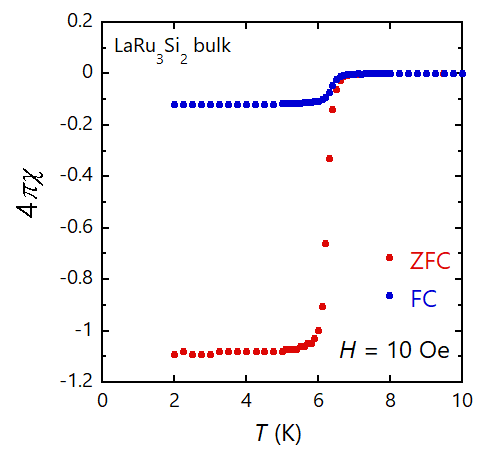}
%\vspace{-2.8cm}
%\includegraphics[width=0.98\linewidth]{diamagneticshift.pdf}
\caption{(Color online) The temperature dependence of zero-field cooled and field-cooled dc magnetic susceptibility of LaRu$_{3}$Si$_{2}$, measured in 1~mT.}
\label{fig7}
\end{figure*}
%%%%%%%%%%%%%%%%%%%%%%%%%%%%%%%%%%%%%%%%%%%%%%%%%%%%

%%%%%%%%%%%%%%%%%%%%%%%%%%%%%%%%%%%%%%%%%%%%%%%%%%%%%%%%%%%%
\begin{figure*}[t!]
\centering
\includegraphics[width=0.7\linewidth]{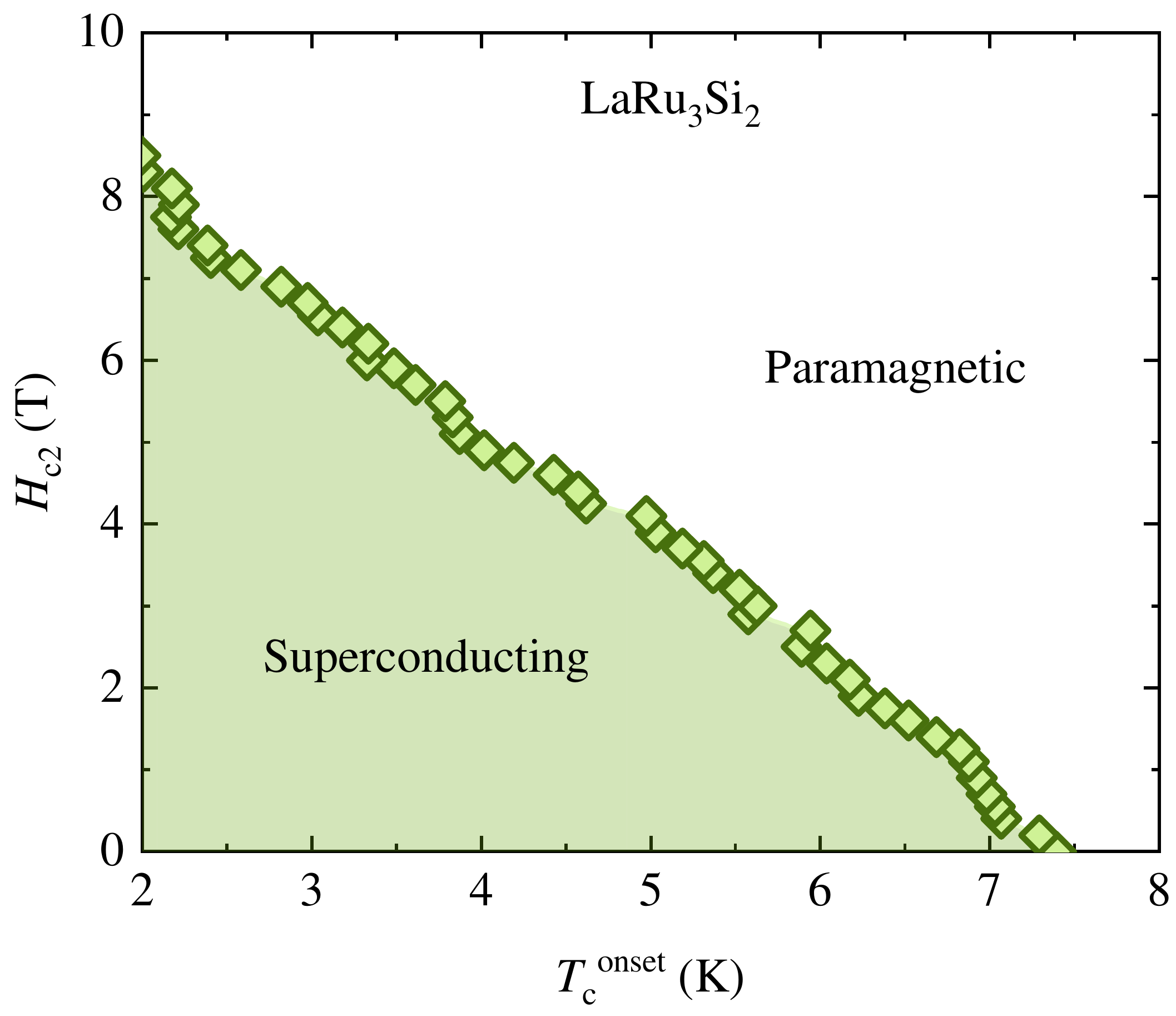}
\vspace{0.0cm}
\caption{ (Color online) The temperature dependence of the second critical magnetic field for LaRu$_{3}$Si$_{2}$.}
\label{fig4}
\end{figure*}
%%%%%%%%%%%%%%%%%%%%%%%%%%%%%%%%%%%%%%%%%%%%%%%%%%%%%%%%%%%%

\textbf{Pressure cell}:  Pressures up to 1.9 GPa were generated in a double wall piston-cylinder
type cell made of CuBe, specially designed to perform ${\mu}$SR experiments under
pressure \cite{GuguchiaPressure,KhasanovPressure}. As a pressure transmitting medium Daphne oil was used. The pressure was measured by tracking the SC transition of a very small indium plate by AC susceptibility. The filling factor of the pressure cell was maximized. The fraction of the muons stopping in the sample was approximately 40 ${\%}$.\\

\textbf{${\mu}$SR experiment}:  In a ${\mu}$SR experiment nearly 100 ${\%}$ spin-polarized muons ${\mu}$$^{+}$
are implanted into the sample one at a time. The positively
charged ${\mu}$$^{+}$ thermalize at interstitial lattice sites, where they
act as magnetic microprobes. In a magnetic material the 
muon spin precesses in the local field $B_{\rm \mu}$ at the
muon site with the Larmor frequency ${\nu}_{\rm \mu}$ = $\gamma_{\rm \mu}$/(2${\pi})$$B_{\rm \mu}$ (muon
gyromagnetic ratio $\gamma_{\rm \mu}$/(2${\pi}$) = 135.5 MHz T$^{-1}$). 
Using the $\mu$SR technique, important length scales of superconductors can be measured, namely the magnetic penetration depth $\lambda$ and the coherence length $\xi$. If a type II superconductor is cooled below $T_{\rm c}$ in an applied magnetic field ranging between the lower ($H_{c1}$) and the upper ($H_{c2}$) critical fields, a vortex lattice is formed which in general is incommensurate with the crystal lattice, with vortex cores separated by much larger distances than those of the crystallographic unit cell. Because the implanted muons stop at given crystallographic sites, they will randomly probe the field distribution of the vortex lattice. Such measurements need to be performed in a field applied perpendicular to the initial muon spin polarization (so-called TF configuration). 

 ${\mu}$SR experiments under pressure were performed at the ${\mu}$E1 beamline of the Paul Scherrer Institute (Villigen, Switzerland using the instrument GPD, where an intense high-energy ($p_{\mu}$ = 100 MeV/c) beam of muons is implanted in the sample through the pressure cell.\\
  
\textbf{Analysis of TF-${\mu}$SR data}: The TF ${\mu}$SR data were analyzed by using the following functional form \cite{Bastian}:
%%%%%%%%%%%%%%%%%%%%%%%%%%%%%%%%%%%%%%%%%%%%%%%%%%%%%%%%%%%%%%
\begin{equation}
\begin{aligned}
P(t)=A_s\exp\Big[-\frac{(\sigma_{sc}^2+\sigma_{nm}^2)t^2}{2}\Big]\cos(\gamma_{\mu}B_{int,s}t+\varphi) \\
 + A_{pc}\exp\Big[-\frac{\sigma_{pc}^2t^2}{2}\Big]\cos(\gamma_{\mu}B_{int,pc}t+\varphi), 
%\label{eq3}
\end{aligned}
\end{equation}
%%%%%%%%%%%%%%%%%%%%%%%%%%%%%%%%%%%%%%%%%%%%%%%%%%%%%%%%%%%%%%
 Here $A_{\rm s}$ and $A_{\rm pc}$  denote the initial assymmetries of the sample and the pressure cell, respectively. 
${\varphi}$ is the initial phase of the muon-spin ensemble and $B_{\rm int}$ represents the
internal magnetic field at the muon site. The relaxation rates ${\sigma}_{\rm sc}$ 
and ${\sigma}_{\rm nm}$ characterize the damping due to the formation of the FLL in the SC state and of the nuclear 
magnetic dipolar contribution, respectively. In the analysis, ${\sigma}_{\rm nm}$ was assumed 
to be constant over the entire temperature range and was fixed to the value obtained above 
$T_{\rm c}$ where only nuclear magnetic moments contribute to the muon depolarization rate ${\sigma}$.
The Gaussian relaxation rate, ${\sigma}_{\rm pc}$, reflects the depolarization due
to the nuclear moments of the pressure cell. 
The width of the pressure cell signal increases below $T_{c}$. As shown previously \cite{GuguchiaNature}, this is due to 
the influence of the diamagnetic moment of the SC sample on the pressure cell, leading to the temperature dependent
${\sigma}_{\rm pc}$ below $T_{c}$. In order to consider this influence we assume the linear coupling between ${\sigma}_{\rm pc}$ and the field shift of the internal magnetic field in the SC state: 
${\sigma}_{\rm pc}$($T$) = ${\sigma}_{\rm pc}$($T$ ${\textgreater}$ $T_{\rm c}$) + $C(T)$(${\mu}_{\rm 0}$$H_{\rm int,NS}$ - ${\mu}_{\rm 0}$$H_{\rm int,SC}$), where  ${\sigma}_{\rm pc}$($T$ ${\textgreater}$ $T_{\rm c}$) = 0.25 ${\mu}$$s^{-1}$ is the temperature independent Gaussian relaxation rate. ${\mu}_{\rm 0}$$H_{\rm int,NS}$ and ${\mu}_{\rm 0}$$H_{\rm int,SC}$ are the internal magnetic fields measured in the normal and in the SC state, respectively. 
As indicated by the solid lines in Figs.~2b,c of the main manuscript the ${\mu}$SR data are well described by Eq.~(1).
The good agreement between the fits and the data demonstrates that the model used 
describes the data rather well.\\

%\textbf{Analysis of ${\lambda}(T)$}:

%As pointed out in the manuscript, for polycrystalline samples the temperature dependence of the London magnetic penetration depth 
%${\lambda}(T)$ is related to the muon spin depolarization rate ${\sigma}_{\rm sc}(T)$ by the Eq. ~1 (see the main text).
%Equation (1) is valid, when the separation between the vortices is smaller than ${\lambda}$ and the applied field small with respect to the second critical field $B_{\rm c2}$. In this case according
%to the London model ${\sigma}_{\rm sc}$ is field independent \cite{Brandt}.
%Field dependent measurements of ${\sigma}_{\rm sc}$ at ambient pressure
%was reported previously \cite{GuguchiaN}. It was observed that first
%${\sigma}_{\rm sc}$ strongly increases with increasing magnetic field until reaching a maximum
%at ${\mu}_{\rm 0}H$  ${\simeq}$  0.03~T and then above 0.03~T stays nearly constant up to the highest field (0.64~T) investigated. Such a behavior is expected within the London model and is typical for polycrystalline high temperature superconductors (HTS's) \cite{Pumpin}.}

% The observed field dependence of ${\sigma}_{\rm sc}$ implies that for a reliable determination of
%the penetration depth the applied field must be larger than ${\mu}_{\rm 0}H= 0.03$~T.}

\textbf{Analysis of ${\lambda}(T)$}: ${\lambda}$($T$) was calculated within the local (London) approximation (${\lambda}$ ${\gg}$ ${\xi}$) by the following expression \cite{Bastian,Tinkham}:
%%%%%%%%%%%%%%%%%%%%%%%%%%%%%%%%%%%%%%%%%%%%%%%%%%%%%%%%%%%%%%%%%%%%%
\begin{equation}
\frac{\lambda^{-2}(T,\Delta_{0})}{\lambda^{-2}(0,\Delta_{0})}=
1+\frac{1}{\pi}\int_{0}^{2\pi}\int_{\Delta(_{T,\phi})}^{\infty}(\frac{\partial f}{\partial E})\frac{EdEd\phi}{\sqrt{E^2-\Delta_i(T,\phi)^2}},
\end{equation}
%%%%%%%%%%%%%%%%%%%%%%%%%%%%%%%%%%%%%%%%%%%%%%%%%%%%%%%%%%%%%%%%%%
where $f=[1+\exp(E/k_{\rm B}T)]^{-1}$ is the Fermi function, ${\phi}$ is the angle along the Fermi surface, and ${\Delta}_{i}(T,{\phi})={\Delta}_{0}{\Gamma}(T/T_{\rm c})g({\phi}$)
(${\Delta}_{0}$ is the maximum gap value at $T=0$). 
The temperature dependence of the gap is approximated by the expression 
${\Gamma}(T/T_{\rm c})=\tanh{\{}1.82[1.018(T_{\rm c}/T-1)]^{0.51}{\}}$,\cite{carrington} 
while $g({\phi}$) describes 
the angular dependence of the gap and it is replaced by 1 for both an $s$-wave and an $s$+$s$-wave gap,
and ${\mid}\cos(2{\theta}){\mid}$ for a $d$-wave gap.\\

%%%%%%%%%%%%%%%%%%%%%%%%%%%%%%%%%%%%%%%%%%%%%%%%%%%%%%%%%%%%%
\begin{figure*}[b!]
\centering
\includegraphics[width=1.2\linewidth]{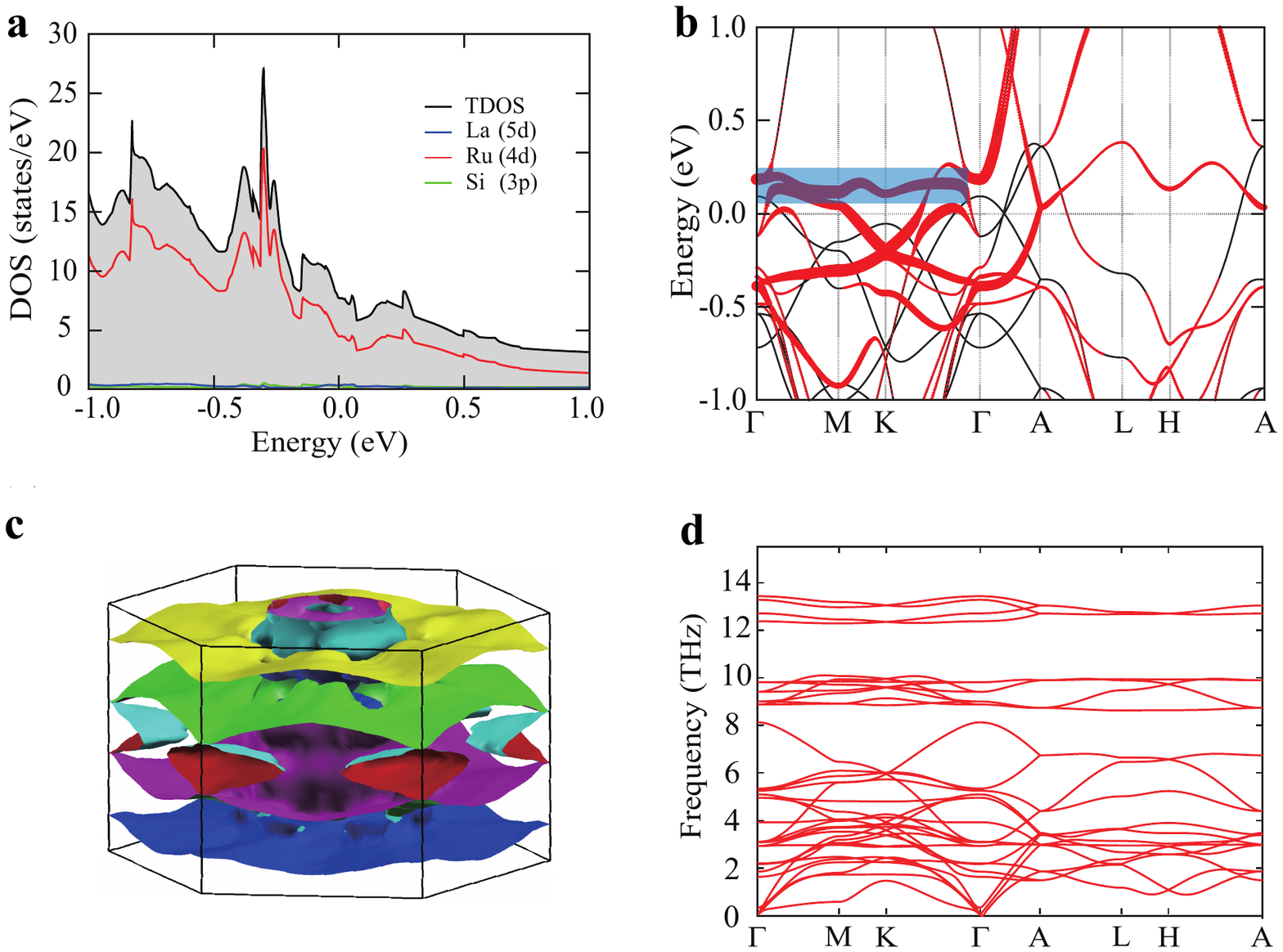}
\vspace{-3.3cm}
\caption{ (Color online)  \textbf{Non-SOC electronic structures and phonon dispersion calculated by GGA+U method with U = 1 eV.} (a) The calculated total DOS and projected DOS for the Ru, Si, La atoms in the bulk LaRu$_{3}$Si$_{2}$. (b)The band structures (black) and orbital-projected band structure (red) for the Ru-$dz^{2}$ orbital without SOC along the high symmetry $k$-path. The width of the line indicates the weight of each component. The blue-colored region highlights the manifestation of the kagome flat band. (c) The Fermi surface of LaRu$_{3}$Si$_{2}$ in the first Brillouin zone. (d) The phonon dispersion in bulk LaRu$_{3}$Si$_{2}$.}
\label{fig4}
\end{figure*}
%%%%%%%%%%%%%%%%%%%%%%%%%%%%%%%%%%%%%%%%%%%%%%%%%%%%%%%%%%%%%%   

\textbf{Electronic structures and phonon dispersion}: The first-principles calculations are performed by the Vienna $ab$ $initio$ simulation package (VASP) \cite{Kresse} based on density functional theory (DFT). The generalized gradient approximation (GGA) of Perdew, Burke, and Ernzerhof (PBE) \cite{Perdew} + Hubbard $U$ \cite{Anisimov} is used for the exchange correlation potential with U = 1 eV for Ru 4$d$ orbitals. The cut-off energy for the wave function expansion is 380 eV, and 12 ${\times}$ 12 ${\times}$ 12 $k$-point sampling grids are used in all calculations. The structure of bulk LaRu$_{3}$Si$_{2}$ is optimized until the force on each atom is less than 0.01 eV/Å, and a 2 ${\times}$ 2 ${\times}$ 2  supercell was built to calculate the phonon dispersion by using the VASP-DFPT(density functional perturbation theory) \cite{Gianozzi} and Phonopy \cite{Togo}. The optimized lattice constants $a$ = $b$ = 5.832 Å, and $c$ = 6.883 Å are used in all calculations.

%%%%%%%%%%%%%%%%%%%%%%%%%%%%%%%%%%%%%%%%%%%%%%%%%%%%%%%%%%%%%
\begin{figure*}[b!]
\centering
\includegraphics[width=1.0\linewidth]{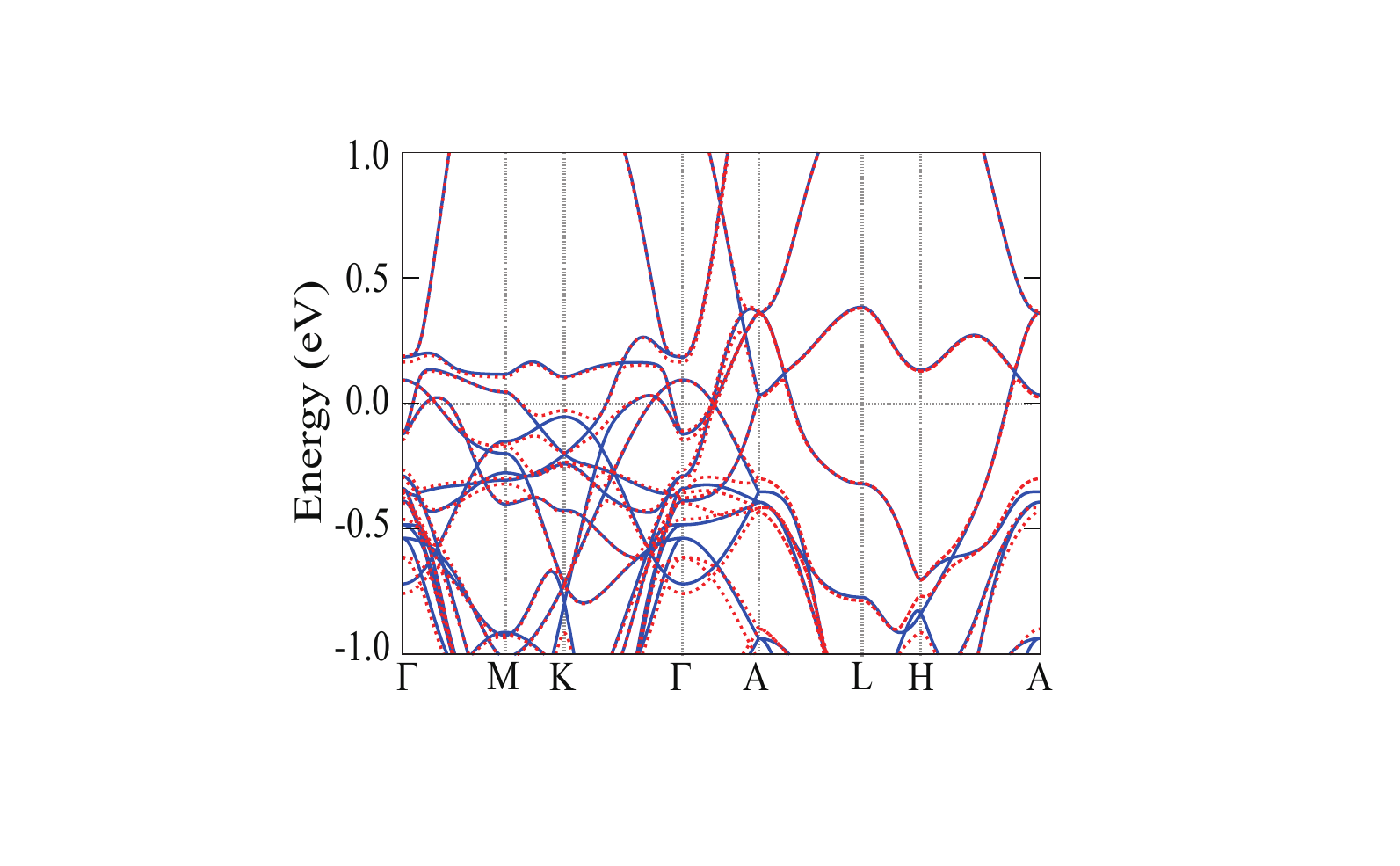}
\vspace{-2.3cm}
\caption{ (Color online)  \textbf{Non-SOC and SOC electronic structure for LaRu$_{3}$Si$_{2}$.}}
\label{fig4}
\end{figure*}
%%%%%%%%%%%%%%%%%%%%%%%%%%%%%%%%%%%%%%%%%%%%%%%%%%%%%%%%%%%%%%   

The calculated total and projected density of states (DOS) of LaRu$_{3}$Si$_{2}$ are presented in supplementary Figure S7a. The DOS at the Fermi level is ${\simeq}$ 7.6 states/eV, corresponding to its experimental metallic behavior. The projected DOS demonstrates that the states at the Fermi level are mainly contributed by the Ru 4$d$ electrons. The calculated band structure with the Ru-$dz^{2}$ orbital projection is shown in supplementary Figure S7b. There are several bands that cross the Fermi level, indicating multi-band physics. The aforementioned single gap superconductivity in this multiband system implies that the SC pairing occurs only on one specific band. The strong dispersion along the ${\Gamma}$-A  line indicates that LaRu$_{3}$Si$_{2}$ is a 3-dimensional material. In the $k_{z}$ = 0 plane, a typical flat band of the kagome lattice formed by the Ru-$dz^{2}$ orbitals is found 0.1 eV above the Fermi level, and a Dirac point at the K (K$^{`}$)-point with linear dispersion is found 0.2 eV below the Fermi level. For several bands crossing the Fermi level, complicated 3-dimensional Fermi surfaces are formed in the first Brillouin zone as shown in Supplementary Figure S7c. All these results are implemented without SOC, because the DOS, band dispersions and also the shape of the Fermi surfaces are affected very little by the SOC effect. We show the band structures with and without SOC for comparison in the supplementary Figure. S8. Some non-trivial band splitting may be present at the Fermi level when SOC is considered, which may bring some novel Berry curvature physics, such as the anomalous Hall effect. 

%%%%%%%%%%%%%%%%%%%%%%%%%%%%%%%%%%%%%%%%%%%%%%%%%%%%%%%%%%%%
\begin{figure*}[t!]
\centering
\includegraphics[width=1.0\linewidth]{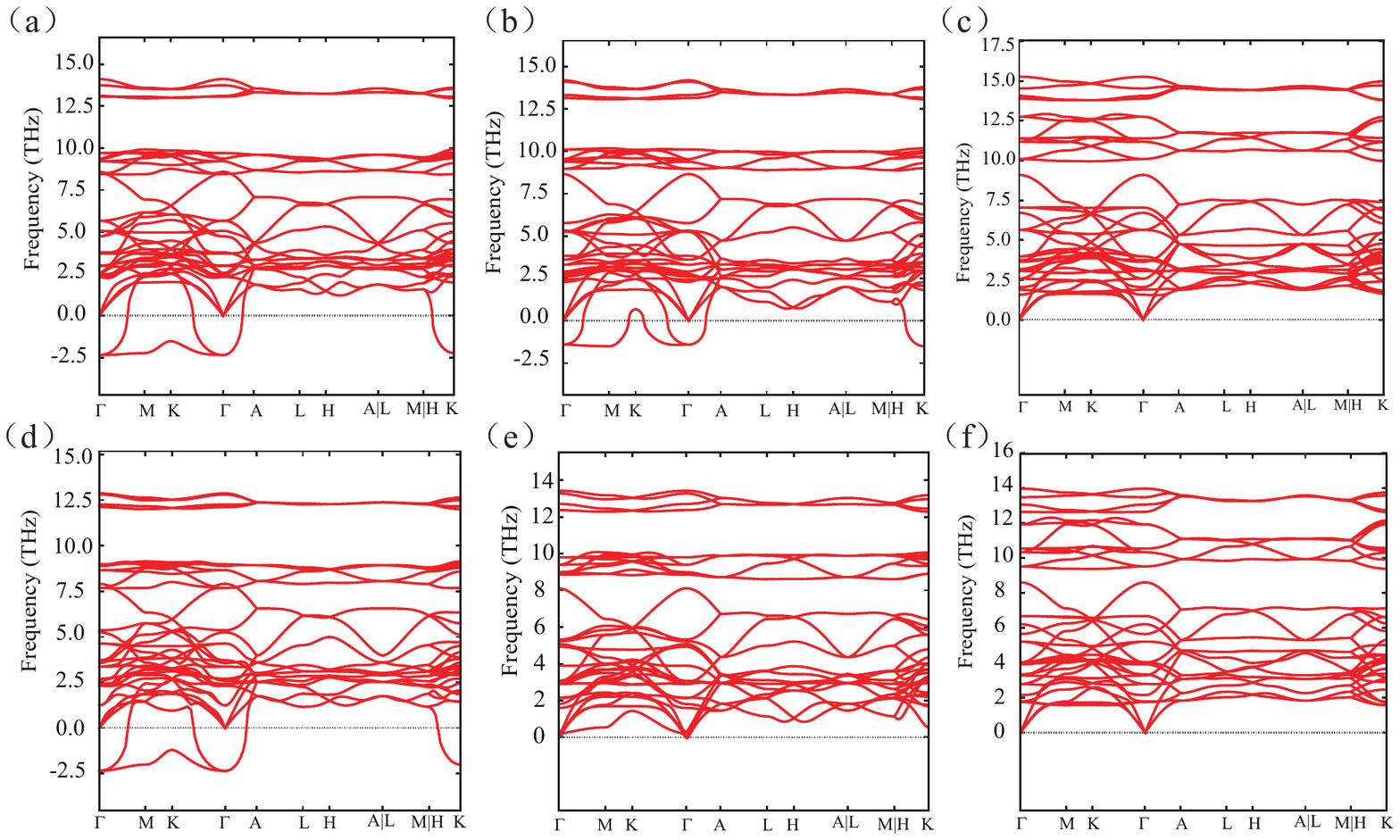}
\vspace{-14.5cm}
\caption{ (Color online)  \textbf{The phonon dispersion in the bulk of LaRu$_{3}$Si$_{2}$.} A phonon dispersion obtained from local density approximation (LDA) + Hubbard $U$ (a-c) and generalized gradient approximation (GGA) + Hubbard $U$ (d-f) frameworks for different values of $U$.}
\label{fig4}
\end{figure*}
%%%%%%%%%%%%%%%%%%%%%%%%%%%%%%%%%%%%%%%%%%%%%%%%%%%%%%%%%%%% 

%%%%%%%%%%%%%%%%%%%
\begin{table*}[t!]
\caption{The optimized lattice constants with the Local Density Approximation (LDA).}
\vspace{0.3cm}
\begin{tabular}{lcccc}
\hline
\hline
                                            &  U=0 eV    & U=1.7 eV & U=1.8 eV \\ \hline 
a(Å)                              &  5.568  & 5.635 & 5.849 \\ 
c(Å)                           &  7.130 & 6.944 & 6.474  \\
%2${\Delta}_{2}/k_{\rm B}T_{\rm c}$       & 5.8(6) & 5.7(5) & 5.1(4) & 4.5(4)\\ 
%%${\omega}_{1}$                               & 0.19(5) & 0.21(4) & 0.15(3) & 0.36(3)\\
%%${\lambda}$ (nm)                       & 249(15) & 250 (17) & 255 (9) & 267(5) \\  \hline
\hline         
\end{tabular}
\label{table1}
\end{table*}
%%%%%%%%%%%%%%%%%%%%%%%%%%%%%%%%%%%%%%%%%%%%%%%%%%%%%%%%%%%%

%%%%%%%%%%%%%%%%%%%
\begin{table*}[t!]
\caption{The optimized lattice constants with the Generalized Gradient Approximation (GGA).}
\vspace{0.3cm}
\begin{tabular}{lcccc}
\hline
\hline
                                            &  U=0 eV    & U=1.0 eV & U=1.1 eV \\ \hline 
a(Å)                              &  5.680  & 5.832 & 5.959 \\ 
c(Å)                           &  7.216 & 6.883 & 6.599  \\
%2${\Delta}_{2}/k_{\rm B}T_{\rm c}$       & 5.8(6) & 5.7(5) & 5.1(4) & 4.5(4)\\ 
%%${\omega}_{1}$                               & 0.19(5) & 0.21(4) & 0.15(3) & 0.36(3)\\
%%${\lambda}$ (nm)                       & 249(15) & 250 (17) & 255 (9) & 267(5) \\  \hline
\hline         
\end{tabular}
\label{table1}
\end{table*}
%%%%%%%%%%%%%%%%%%%%%%%%%%%%%%%%%%%%%%%%%%%%%%%%%%%%%%%%%%%%

To confirm the dynamic stability of LaRu$_{3}$Si$_{2}$, the phonon dispersion has been calculated. We find that the lattice constant and phonon dispersion is very sensitive to the exchange correlation potential and the value of the Hubbard $U$. A phonon dispersion with negative frequency modes is obtained when only the GGA or LDA type of exchange correlation potential is used (See supplementary Fig. S9a-f). In the GGA + Hubbard $U$ framework, the negative frequency modes can be eliminated when U ${\textgreater}$ 1 eV. However, the lattice constant $c$ shows a sudden collapse when U increases minutely. For example, $c$ = 6.604 Å is obtained at $U$ = 1.1 eV, which is lower by 7 ${\%}$ than the experimental parameter. Moreover, we notice that the electronic structures shown in the supplementary Fig. S7 are not sensitive to the exchange correlation functional.  All these results indicate that Coulomb repulsion $U$ plays a crucial role in stabilizing the crystal structure of LaRu$_{3}$Si$_{2}$, indicating the medium strength correlations in this system. This is reminiscent of another family of correlated metals, the parent compounds of iron-based superconductors, in which GGA/LDA is also unable to describe bond length and bond strength, but can describe band structures quantitively well. 

 The phonon dispersion for LaRu$_{3}$Si$_{2}$, calculated by the GGA+U method with U = 1 eV, is shown in supplementary Figure S7d. The cut-off energy for the wave function expansion is 380 eV. It consists of only the positive frequency modes, and the optimized lattice constants agree very well with the experimental parameters. Based on these parameters, we further calculated the electron-phonon coupling constant ${\lambda}$ to be ${\sim}$ 0.45. By using the Debye temperature ${\theta}$ ${\simeq}$ 280 K, ${\mu}^{*}$ = 0.12, the superconducting transition temperature $T_{\rm c}$ induced by the electron-phonon coupling can be estimated to be ${\sim}$ 1 K, which is obviously smaller than the experimental result. Since the electron-phonon coupling can reproduce only  some fraction of the experimental $T_{\rm c}$, other sources which enhance $T_{\rm c}$ must be considered, which is discussed in the main text.

 %Since the calculations show the presence of a flat band relatively close to the Fermi level, we speculate that it might lead to further enhancement of $T_{\rm c}$ in this system.

\bibliographystyle{prsty}

\end{document}